\documentclass[sigplan,screen,authorversion,nonacm]{acmart}

\usepackage[utf8]{inputenc}
\usepackage[T1]{fontenc}

\usepackage{graphicx}
\usepackage[frozencache]{minted}
\usepackage{subcaption}
\usepackage{xcolor}
\usepackage{algorithm}
\usepackage{amsmath}
\usepackage{tabularx}
\usepackage{booktabs}
\usepackage{amsthm}
\usepackage{enumitem}

\definecolor{babcolor}{rgb}{0.9,0.45,0.1}
\definecolor{kaycolor}{rgb}{0.06,0.07,0.62}
\definecolor{abcolor}{rgb}{0.62,0.15,0.06}

\newcommand{\edit}{}

\newcommand{\beginminted}{\begin{minted}[fontsize=\footnotesize]{python}}

\copyrightyear{2024}
\acmYear{2024}
\setcopyright{rightsretained}
\acmConference[ICS '24]{Proceedings of the 38th ACM International Conference on Supercomputing}{June 4--7, 2024}{Kyoto, Japan}
\acmBooktitle{Proceedings of the 38th ACM International Conference on Supercomputing (ICS '24), June 4--7, 2024, Kyoto, Japan}\acmDOI{10.1145/3650200.3656623}
\acmISBN{979-8-4007-0610-3/24/06}

\begin{document}
\title{RDMA-Based Algorithms for Sparse Matrix Multiplication on GPUs}

\author{Benjamin Brock}
\authornote{This work was conducted while Benjamin Brock was a PhD student at the University of California, Berkeley.}
\affiliation{
  \institution{Intel Corporation}
  \city{Santa Clara}
  \state{CA}
  \country{USA}
}
\email{benjamin.brock@intel.com}

\author{Ayd{\i}n Bulu\c{c}}
\affiliation{
  \institution{University of California, Berkeley}
  \city{Berkeley}
  \state{CA}
  \country{USA}
}
\email{abuluc@lbl.gov}

\author{Katherine Yelick}
\affiliation{
  \institution{University of California, Berkeley}
  \city{Berkeley}
  \state{CA}
  \country{USA}
}
\email{yelick@cs.berkeley.edu}

%\author{\IEEEauthorblockN{
%    Benjamin Brock\IEEEauthorrefmark{1},
%    Ayd{\i}n Bulu\c{c}\IEEEauthorrefmark{2}\IEEEauthorrefmark{1},
%    Katherine Yelick\IEEEauthorrefmark{1}\IEEEauthorrefmark{2},}  \\
%  \IEEEauthorblockA{\IEEEauthorrefmark{1}
%    EECS Department,
%    University of California, Berkeley, CA }
%      \IEEEauthorblockA{\IEEEauthorrefmark{2}
%    Computational Research Department,
%    Lawrence Berkeley National Laboratory, Berkeley, CA }
%}

%\email{webmaster@marysville-ohio.com}
%\affiliation{%
%  \institution{Institute for Clarity in Documentation}
%  \streetaddress{P.O. Box 1212}

% \renewcommand{\shortauthors}{}


%%
%% end of the preamble, start of the body of the document source.

\begin{abstract}
Sparse matrix multiplication is an important kernel for large-scale graph processing and other data-intensive applications. In this paper, we implement various asynchronous, RDMA-based sparse times dense (SpMM) and sparse times sparse (SpGEMM) algorithms, evaluating their performance running in a distributed memory setting on GPUs. Our RDMA-based implementations use the NVSHMEM communication library for direct, asynchronous one-sided communication between GPUs. We compare our asynchronous implementations to state-of-the-art bulk synchronous GPU libraries as well as a CUDA-Aware MPI implementation of the SUMMA algorithm. We find that asynchronous RDMA-based implementations are able to offer favorable performance compared to bulk synchronous implementations, while also allowing for the straightforward implementation of novel work stealing algorithms.
\end{abstract}

\maketitle

%\begin{CCSXML}
%\end{CCSXML}

%\ccsdesc[500]{Computer systems organization~Embedded systems}

% TODO: keywords, categories

%\keywords{datasets, neural networks, gaze detection, text tagging}


%\bab{Ben can write comments like this.}
%\kay{Kathy can write comments like this.}
%\ab{Ayd{\i}n can write comments like this.  Overleaf also has a system for adding comments, but those comments won't show up on the Git repo.}

\section{Introduction}
\label{sec:intro}

Sparse matrix multiplication is an important computational primitive that arises in simulation, data analysis, and machine learning applications. Typically limited by memory and network performance, these computations are especially challenging for unstructured matrices, such as those occurring in graph neural networks, genomics, graph analytics, and other data intensive problems.  Sparse matrix  primitives provide a convenient and important set of operations for performance tuning that will impact a wide array of applications, and there is a large body of prior work  optimizing sparse matrix kernels for multicore~\cite{patwary2015parallel}, manycore~\cite{saule2013performance},
GPU~\cite{yang2018design},
%\cite{ortega2014fastspmm,yang2018design},
and distributed memory environments~\cite{schubert2011hybrid,acer2016improving,hong2019adaptive,solomonik2015sparse,gu2020bandwidth}.  Two of the most important kernels are sparse times dense matrix multiplication (SpMM), where the dense matrix is tall and skinny, i.e., representing a set of vectors, and sparse times sparse matrix multiplication (SpGEMM).  Sparse matrix-vector multiplication is also an important kernel in many applications, and has been studied even more extensively, although our focus here is on the matrix-matrix operations.

% TODO cite SUMMA

The large majority of prior work on distributed-memory sparse matrix kernels focuses on bulk synchronous implementations based on SUMMA~\cite{van1997summa}, using collective operations, typically broadcast and/or reduce, in the inner loop of the algorithm to move data to the processes that need it for their local matrix multiplications.  While SUMMA-like algorithms have the advantages of being straightforward to implement and generally scaling well, for sparse problems, they have the disadvantage of forcing processes to proceed in lockstep in the inner loop of the algorithm.  When there is load imbalance, either in the amount of computation that must be performed locally by each process, the amount of data that must be transferred to each process, or both, this can potentially result in decreased performance. This can occur even when the total amount of work performed by each process is the same, since processes may have differing amounts of work in each iteration. To illustrate,  Figure~\ref{fig:perstagelb} shows two kinds of load balance for squaring a randomized sparse matrix generated by the R-MAT model~\cite{chakrabarti2004r}, with parameters $a=0.6$, $b=c=d=0.4/3$,  edgefactor $8$, and scale $17$. The total (end-to-end) computation has only $20\%$ load imbalance whereas the synchronization points in between stages amplify the load imbalance to $\approx 2.3 \times$. The load imbalance is defined as the max/avg ratio, the ratio of maximum number of flops performed by any processor to the average number of flops per processor.

Load imbalance is often addressed by randomly permuting the rows and/or columns of the sparse matrices, but this has disadvantages, the foremost being that the required permutation of the inputs, followed by permutation of the outputs, can be expensive. Furthermore, permuting the matrix may possibly remove structural locality, resulting in decreased performance of local matrix multiply operations.
For example, Slota et al.~\cite{slota2017order} showed that the loss of locality due to random permutations can hurt the performance of graph algorithms significantly where parallel PageRank performs $2-5\times$ slower on a randomly permuted graph compared to other orderings. Even if we ignore the loss of locality or try to recover it by local reordering after a global random permutation, there are theoretical limits on how much load balance random permutations can achieve. For matrices with dense rows or columns, the load balance can deteriorate with $\sqrt{p}$ for large $p$, where $p$ is the number of processors in the worst case, per Theorem 5.4 of Azad et al.~\cite{azad2020distributed}. 

%Finally, perhaps most importantly, synchronous implementations of popular algorithms such as SUMMA and Cannon~\cite{cannon1969cellular}, where processors synchronize in between stages, artificially introduce load imbalance even when the end-to-end load imbalance among processors is negligible. For instance, 
\begin{figure}[thb]
 \centering
 \subcaptionbox{End-to-end load imbalance when processors do not synchronize across $16$ stages. The heat plot has $256 = 16 \times 16$ boxes, which demonstrate the max/avg load imbalance is $\approx 1.2$.}
 {%
 \includegraphics[width=.48\linewidth]{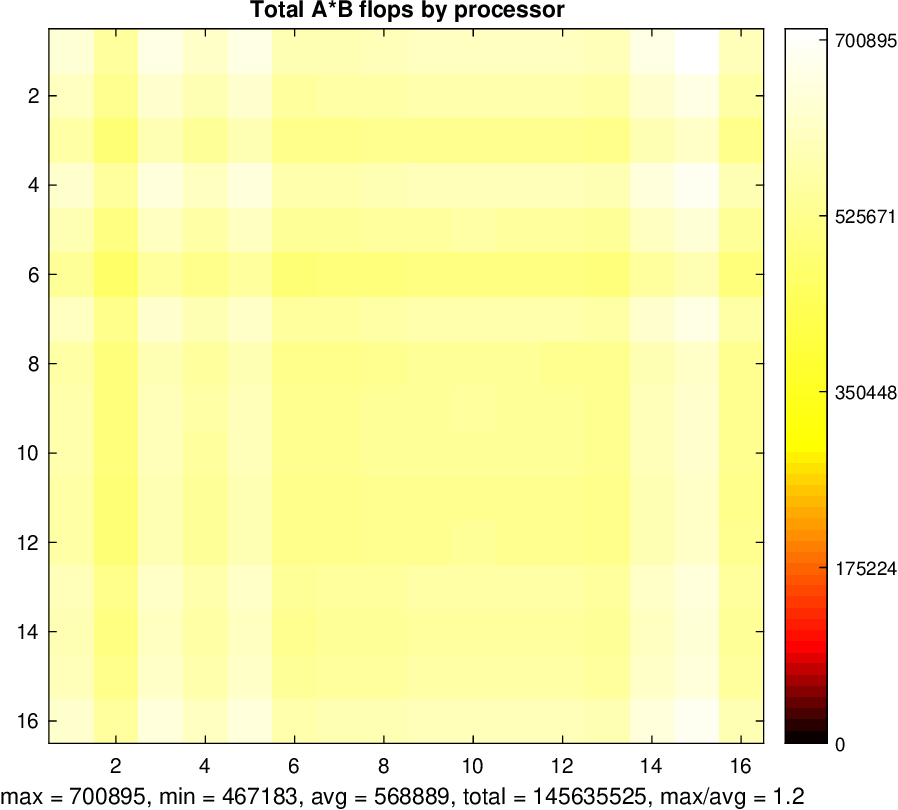}%
 }
 \hfill
 \subcaptionbox{Per stage max/avg load imbalance of the same algorithm. An implementation that synchronizes in between stages would consequently incur load imbalance of $\approx 2.3$.}
 {%
 \includegraphics[width=.48\linewidth]{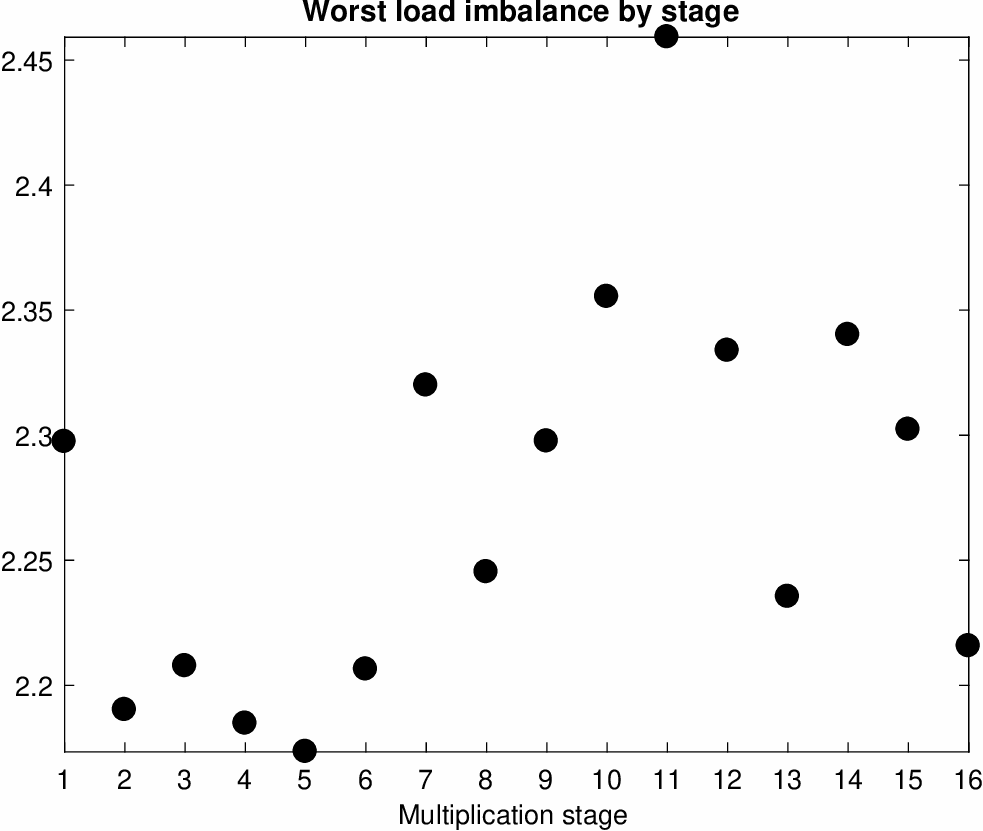}%
 }
 \caption{Total (end-to-end) vs.\, per-stage load balance multiplying a R-MAT model-generated sparse matrix with a sparse 2D algorithm. Simulated on a $16 \times 16$ process grid.}
 \label{fig:perstagelb}
 \end{figure}

\textit{Asynchronous} algorithms, which do not require synchronization or coordination between processes as in bulk synchronous algorithms, are an approach to dealing with this load imbalance in distributed sparse matrix multiplication. In this paper, we develop dense and sparse matrix data structures that use RDMA, meaning that a process can manipulate remote parts of the matrix using one-sided RDMA put and get operations, without requiring synchronization or coordination with the remote process.  Using these dense and sparse matrix data structures, we then design and implement a number of different \textit{RDMA-based, asynchronous} matrix multiply algorithms, which remove synchronization from the inner loop, allowing each process to proceed independently.  In addition, we develop and implement an RDMA-based workstealing algorithm, which allows processes that have finished early to steal work from processes that are overburdened using a lightweight reservation scheme.

Furthermore, we examine the performance of these different sparse matrix multiply algorithms running across multiple GPUs in a distributed memory environment.  We examine the challenges presented by the highly compute dense nodes of modern GPU-based supercomputers, as well as the benefits of direct GPU-to-GPU transfers offered by new network technologies like NVIDIA GPUDirect RDMA and NVLink.

The main contributions of this paper are:

\begin{itemize}[topsep=0pt,itemsep=-1ex,partopsep=1ex,parsep=1ex,leftmargin=*]
  \item Design and implementation of RDMA-based distributed dense and sparse matrix data structures for GPUs.
  \item New asynchronous, RDMA-based algorithms for SpMM and SpGEMM, including workstealing algorithms.
  \item A roofline-based performance model, to characterize the performance of our RDMA-based implementations.
  \item An in-depth performance analysis of our RDMA-based implementations, along with a comparison to traditional bulk synchronous methods.
\end{itemize}

\section{Background}
\label{sec:background}

Sparse times dense matrix multiply (SpMM) and sparse times sparse matrix multiply (SpGEMM) are two important sparse linear algebra primitives.
SpMM is used in a variety of blocked iterative methods, graph algorithms, as well as graph neural networks (GNNs)~\cite{huang2020ge,tripathy2020reducing,hu2020featgraph}. In these contexts, SpMM typically involves multiplication of a sparse matrix with a tall and skinny dense matrix, where the number of columns of the sparse matrix varies between 32 and 1024.
SpGEMM is also used in graph algorithms, including betweenness centrality, triangle counting, cycle detection, and Markov clustering, as well as in applications such as in genomics~\cite{bulucc2011combinatorial,azad2015parallel,dongen2000graph,bustamam2012fast,davis2018graph}. 
% ~\cite{bulucc2011combinatorial,azad2015parallel,yuster2004detecting,dongen2000graph,bustamam2012fast,davis2018graph}. 
% TODO: citation needed for genomics.
% TODO: flesh out with maybe another paragraph about SpMM/SpGEMM.

\subsection{Distributed Matrix Algorithms}
Distributed matrix algorithms allow multiple processes to work together to execute a single matrix multiply operation.  Let's first consider a distributed matrix multiply computing the product $C = AB$.  Here, each of the matrices $C$, $A$, and $B$ is split up into tiles which live on different processes.  The matrix tiles will be sparse or dense, depending on what kind of matrix multiplication is being performed.  Let's assume that $C$ is distributed among a grid of $M \times N$ tiles, $A$ is distributed among a grid of $M \times K$ tiles, and $B$ is distributed among a grid of $K \times N$ tiles. The algorithms we consider in this paper rely on a tiling scheme that is regular on the indices. In other words, each tile has the same dimensions, modulo differences due to the tile size not dividing the matrix size evenly. A different, more irregular tiling mechanism might be more appropriate for matrices from certain applications such as those involving linear-scaling quantum-chemical calculations~\cite{borvstnik2014sparse}, where the near-sightedness creates natural block-sparse structure. However, the vast majority of application targets (GNNs, graph algorithms, eigenvector computations, and tensor decompositions) do not have this natural block-sparse structure, so the added complexity of irregular tiling is not justified for us. 

To compute a single tile of the C output matrix $C_{i,j}$, we must compute the result $C_{i,j} = C_{i,j} + \sum_{k=0}^{K} A_{i,k} B_{k,j}$, where the indices $i$, $j$, and $k$ index into tiles of the matrices. The computation can be thought as a 3D cube, which can be assigned to processors in different ways, which result in 1D, 2D, and 3D structured algorithms. We are then left with the choice of how to assign pieces of the computation to processors, as well as how to communicate tiles of the matrices necessary to execute the matrix multiply.  One of the most common, and usually easiest to implement, distribution methods is the \textit{stationary C} method, in which the process that owns each block of $C$ will be responsible for executing each of the local matrix multiplications necessary to compute the result for that block.  This means that the A and B matrices will be communicated, while tiles of the C matrix will be available locally.  For each tile of the matrix $C_{i,j}$, the process that owns that tile will be responsible for calculating the output block $C_{i,j} = C_{i,j} + \sum_{k=0}^{K} A_{i,k} B_{k,j}$.

Similarly, we could organize the computation so that the owner of each tile of $A$ will be responsible for executing each of the local matrix multiplication operations in which it is used.  In this \textit{stationary A} method, tiles of $A$ will thus be available locally, while tiles of $B$ will be fetched through some method, and output updates to $C$ will  be sent and accumulated at their final location.  This means that for each tile of A $A_{i,k}$, the process that owns that tile will be responsible for executing $C_{i,j} = C_{i,j} + A_{i,k} B_{k,j}$ for all $j$.  Note that this means that the tile of B $B_{k,j}$ must be somehow retrieved from remote memory, while the result of each local matrix multiply $C_{i,j}$ must be sent and accumulated to the corresponding block of C.  The \textit{stationary B} method is similar, except that $B$ remains stationary, while $A$ must be communicated.

% TODO: should we mention that in stationary A you produce updates for a row-block of A, B a column-block of B?

\subsection{Bulk Synchronous SUMMA}
In bulk-synchronous SUMMA, collective broadcast operations are used to communicate tiles of the matrix. In SUMMA, communicators are created for each tile row of the $A$ matrix and each tile column of the $B$ matrix.  In each iteration of the algorithm, a block of the $A$ matrix is broadcast in each row communicator and a block of the $B$ matrix is broadcast in each column communicator. Each process will compute its corresponding local matrix multiplication, accumulating into its local block of $C$.  Stationary A (or B) algorithms are also possible with SUMMA, by replacing one of the broadcasts with a collective reduction operation to accumulate each block of $C$ into the correct place.  %Pseudocode for SUMMA is shown in Algorithm~\ref{fig:bs_summa}.
% TODO: cite SUMMA

% \begin{algorithm}
% \beginminted
% i,j = C.my_block()
% for k in 0..K-1:
%   # Broadcast k'th tile of A in row
%   broadcast(local_a, k, A.row_comm(i))
%   # Broadcast k'th tile of B in col
%   broadcast(local_b, k, B.col_comm(j))
%   # C is local
%   local_c = C.tile_ref(i, j)
%   local_c += local_a*local_b
% \end{minted}
% \caption{Bulk-synchronous SUMMA implementation.}
% \label{fig:bs_summa}
% \end{algorithm}

As discussed in Section~\ref{sec:intro}, variation in the number of nonzeros, as well as the sparsity patterns, of individual tiles of sparse input matrices can create load imbalance in computation due to the differing numbers of flops that must be performed in each local matrix multiply, as well as load imbalance in communication, due to the different amounts of data that must be transferred by each process. Bulk synchronous SUMMA-like algorithms can suffer due to these load imbalances. 

\subsection{RDMA and Asynchrony}
\label{sec:rdma}
Remote Direct Memory Access (RDMA) is a hardware capability offered by the Network Interface Cards (NICs) in most modern supercomputers and datacenters.  Using RDMA, a process can expose a region of its memory, often referred to as ``shared memory,'' to be directly accessible by remote processes.  Processes can then issue put, get, and atomic operations to remotely access and manipulate the shared memory regions of other processes.  {\edit These RDMA operations are handled directly by the NIC on the remote node, hence no coordination is required in order to access the memory of a remote process}.  Some of the commonly used frameworks that support direct RDMA access include OpenSHMEM, GASNet-EX, and MPI's one-sided communication API.  For distributed programs that run on GPUs, systems with Infiniband can currently take advantage of GPUDirect RDMA, which allows direct transfers from GPU-to-GPU over the network.  NVSHMEM, an extension of OpenSHMEM from NVIDIA, allows for direct transfers of data between GPUs, either within a node over PCIe or NVLink, or between nodes using GPUDirect RDMA over Infiniband. In this paper, we use NVSHMEM to directly copy between GPUs.
% TODO: cite GASNet, OpenSHMEM, MPI, NVSHMEM
A number of different libraries have been developed that use RDMA primitives to build data structures and algorithms, with some focusing particularly on irregular data structures and algorithms, where asynchronous RDMA primitives can be particularly beneficial, both for performance and software engineering reasons~\cite{brock2019bcl,bachan2017upc++,Fuerlinger:2016:DASH}.

% \vspace{-0.5em}
\section{RDMA-Based Algorithms}
% \vspace{-0.5em}
\subsection{Data Structures}
\label{sec:data_structures}
Unlike bulk synchronous SUMMA algorithms, RDMA-based algorithms use direct RDMA put and get operations to access remote data.  In order to implement RDMA-based algorithms, we first developed RDMA-based dense and sparse matrix data structures to support direct remote access to tiles.  In our data structures, each process holds a directory of \textit{global pointers}, which are objects that reference remote data that can be accessed over RDMA.  In the dense case, each process holds a global pointer for each remote tile, and it can issue put or get operations to read or write to the remote matrix tile.  In the sparse matrix data structure, the directory holds three global pointers for each tile, which point to the values, row pointer, and column indices arrays that constitute a Compressed Sparse Row (CSR) format sparse matrix.

% TODO diagram?

\subsubsection{Reading Tiles}
\label{sec:reading_tiles}
To fetch a remote tile of the matrix is fairly straightforward.  For a dense matrix, we issue a single RDMA get operation to copy the data from the remote node to the process's local memory.  For a sparse matrix, we issue three get operations to retrieve each of the remote CSR arrays.  Note that since we use RDMA, these operations are completely asynchronous, and do not require coordination with the remote process.  We implement both blocking and non-blocking versions of the operation.

\subsubsection{Modifying Remote Tiles}
\label{sec:modify_remote}
Modifying a local tile is straightforward, since we can use local pointers which point to a local tile of the matrix to directly modify the data.  Modifying a remote tile, which is only required for some of the RDMA algorithms, is somewhat more complex, particularly when the remote tile happens to be sparse.  To deal with this, we use a system of remote queues to issue asynchronous updates to the remote tiles.  Each process has a globally visible queue, and other processes can push updates to this queue in order to send updates that the remote process needs to apply to tiles that it owns.  During algorithms that require accumulations to remote tiles, each process will periodically check to see if there are elements waiting in the queue, and if so, dequeue the elements and accumulate the tile.  The element inserted into the queue is a lightweight pointer to the submatrix that needs to be accumulated, so the dequeueing process will issue get operations to retrieve the data before accumulating.  Push and pop operations are performed atomically to ensure no updates are lost.

% \vspace{-0.5em}
\subsection{Algorithms}
\subsubsection{RDMA Stationary C Algorithm}
The most straightforward RDMA-based algorithm is the stationary C algorithm, in which each processor iterates through its tiles of the C matrix, and, for each tile, iterates through the corresponding row block of A and column block of B, retrieving each of the remote tiles via \texttt{get\_tile} and multiplying them together using a local matrix multiply operation. 
%Pseudocode for a basic implementation of this algorithm is shown in Algorithm~\ref{fig:rdma_csimple}. Note that 
Since the method \texttt{get\_tile} does not require any coordination with remote processors, each process can execute its work independently, and does not need to wait on or synchronize with other processors. 
%Note that the pseudocode includes a barrier statement at the end, but unless the entire result is required immediately, processes in a real-world program would actually be free to exit the distributed matrix multiply and proceed to do other work.
In order to achieve good performance, a few optimizations are required, including prefetching tiles for the next iteration, which allows overlap of computation and communication, and an iteration offset in the inner loop, which balances which tile is requested within each row and column and generally ensures the first RDMA get is to the local tile.  These optimizations are discussed in detail in Section~\ref{sec:optimizations}.

% \begin{algorithm}
% \beginminted
% for i in 0..M-1:
%   for j in 0..N-1:
%     if C.owner(i, j) == rank():
%       for k in 0..K-1:
%         local_a = A.get_tile(i, k)
%         local_b = B.get_tile(k, j)
%         local_c = C.tile_ref(i, j)
%         local_c += local_a*local_b
% barrier()
% \end{minted}
% \caption{RDMA-based 2D $C$-Stationary SpMM.  M, N, and K represent tile dimensions.}
% \label{fig:rdma_csimple}
% \end{algorithm}

\setlength{\textfloatsep}{0pt}
\begin{algorithm}
\beginminted
for i in 0..M-1:
  for k in 0..K-1:
    if A.owner(i, k) == rank():
      for j in 0..N-1:
        local_a = A.tile_ref(i, k)
        local_b = B.get_tile(k, j)
        local_c = local_a*local_b
        queue[C.owner(i, j)].push(get_ptr(local_c))

while local_c_ptr = queue[rank()].pop():
  C.my_tile() += get_tile(local_c_ptr)
barrier()
\end{minted}
\caption{RDMA-based 2D $A$-Stationary SpMM.  M, N, and K represent tile dimensions.}
\label{fig:rdma_asimple}
\end{algorithm}

% \vspace{-0.25em}
\subsubsection{RDMA Stationary A and B Algorithms}
The RDMA-based stationary A algorithm is similar to the stationary C algorithm, except we assign work based on which processor owns which tile of A.  Each process iterates through its tiles of A, and, for each tile, iterates through the corresponding tile row of B, pulling in each tile and performing a local matrix multiply.  Each of these partial results must be accumulated into different tiles of C, each of which is potentially owned by a different process.  In order to accumulate these partial results into their individual tiles of C, a global pointer to the local result is pushed to a queue on the remote process where it needs to be accumulated.  At some point, the remote process will pop the pointer off of its queue, retrieve the remote partial result tile, and accumulate it into the correct local tile.  Pseudocode for the RDMA-based stationary A algorithm is shown in Alg~\ref{fig:rdma_asimple}.  The stationary B algorithm is very similar, except work is assigned based on which processes own which tile of B.

% Cut this paragraph?
%The stationary B algorithm is very similar, except work is assigned based on which processes own which tile of B.  Each process will iterate through its tiles of B, and for each tile of B, iterate through the corresponding tile column of A, retrieving the tiles of A, performing a local matrix multiply, and pushing a pointer to the partial result onto the remote queue of the process that owns the tile of C to which the partial result must be accumulated.

% \vspace{-0.25em}
\subsection{Optimizations}
\label{sec:optimizations}
Two important optimizations are necessary in order to achieve good performance for the RDMA-based algorithms described here.  First, we use the non-blocking version of \texttt{get\_tile} to asynchronously retrieve a remote tile of the matrix.  This allows us to prefetch the tiles needed for the next iteration before entering into the local matrix multiply operation, creating overlap of communication with computation.  Second, we apply an iteration offset to the inner loop, which ensures two things: (1) it spaces processes apart, so that not all processes will request the same tile in their row and column at the same time, and (2) it generally ensures that the first remote get issued is to a local tile, which helps jumpstart communication and ensures that almost all communication can be overlapped with computation.  For the stationary C algorithm, we apply an iteration offset of $i + j$, while in the stationary A and stationary B algorithms, we apply an iteration offset of $i + k$ and $k + j$, respectively.  Pseudocode which demonstrates these optimizations applied to a stationary C algorithm is shown in Alg~\ref{fig:rdma_copt}.

\begin{algorithm}
\beginminted
for i in 0..M-1:
  for j in 0..N-1:
    if C.owner(i, j) == rank():
      k_offset = i + j
      buf_a = A.async_get_tile(i, k_offset % K)
      buf_b = B.async_get_tile(k_offset % K, j)
      for k_ in 0..K-1:
        k = (k_ + k_offset) % K
        local_a = buf_a.get()
        local_b = buf_b.get()
        local_c = C.tile_ref(i, j)

        if (k_ + 1 < K):
          buf_a = A.async_get_tile(i, (k+1) % K)
          buf_b = B.async_get_tile((k+1) % K, j)
        local_c += local_a*local_b
barrier()
\end{minted}
\caption{Optimized RDMA-based 2D $C$-Stationary SpMM. M, N, and K represent tile dimensions.}
\label{fig:rdma_copt}
\end{algorithm}

% \vspace{-0.25em}
\subsection{Workstealing Algorithms}
In addition to supporting generally asynchronous implementations of distributed matrix multiply, RDMA also allows for more straightforward implementations of workstealing algorithms, in which processes that have finished their work can steal work from processes that are overburdened.  In our workstealing algorithms, we use lightweight 2D and 3D reservation schemes to allow processors to claim work.  First, processors attempt to perform their work as normal, iterating through each of the tiles that they own.  However, before performing any work, they first issue a remote fetch-and-add operation, executed using RDMA, to reserve the work.  If some of a process' work is stolen, it will move on to the next piece of work available.  After it has completed all of its normal work, each process will iterate through a number of other processes, checking to see if they have work available to be stolen.

\paragraph{Random workstealing}
The simplest workstealing strategy is random workstealing, where work is stolen randomly without regard to locality.  This has the advantage that many blocks are available to be stolen, with the disadvantage that performing stolen work will usually be more expensive than performing regular work, since the tiles of A, B, and C must all be communicated, as opposed to just two.  Random workstealing uses a 2D work grid corresponding to the tiles of the stationary matrix.  Each element in the work grid contains an integer that is initialized to zero.  To claim a piece of work involving a tile, processes perform a fetch-and-add operation on the corresponding element of the work grid.  The integer value returned corresponds to the piece of work that has been claimed.
Pseudocode for this workstealing algorithm is shown in Alg~\ref{fig:rdma_ws} using the stationary A method.

\paragraph{Locality-Aware workstealing}
In locality-aware workstealing, instead of stealing randomly, each process will only steal pieces of work for which it owns one of the components, ensuring a lower cost of performing stolen work.
For example, in a stationary C locality-aware work-stealing algorithm, processes will first attempt to perform matrix multiplies involving their tiles of C.  After these are completed, each process will attempt to steal work involving their tiles of A and B.
Locality-aware workstealing requires a 3D work grid, where element $i,j,k$ corresponds to the component matrix multiply $C[i, j] += A[i, k]*B[k, j]$.  There is a slightly higher overhead associated with this 3D work grid, which requires a separate remote fetch-and-add for each piece of work performed, but this is largely offset by the improvement in locality.

\begin{algorithm}
\beginminted
def attempt_work(i, k):
  # Remote atomic fetch-and-add to reserve work
  my_j = reserve_grid[i, k]++
  while my_j < N:
    local_a = A.get_tile(i, k)
    local_b = B.get_tile(k, j)
    local_c = local_a*local_b
    
    queue[C.owner(i, j)].push(get_ptr(local_c))
    my_j = reserve_grid[i, k]++

# Do work for my tiles
for i in 0..M-1:
  for k in 0..K-1:
    if A.owner(i, k) == rank():
      attempt_work(i, k)

# Attempt to steal work
for idx in 0..M-1:
  i = (rank()+idx) / M
  k = (rank()+idx) % M
  attempt_work(i, k)

while local_c_ptr = queue[rank()].pop():
  C.my_tile() += get_tile(local_c_ptr)
barrier()
\end{minted}
\caption{Stationary A SpMM with random workstealing.}
\label{fig:rdma_ws}
\end{algorithm}

% \vspace{-0.5em}
\section{Performance Model}
\label{sec:perf_model}
In this section, we build a performance model to characterize and predict the performance of our RDMA-based matrix multiply algorithms, first computing the expected cost of communication before using this communication analysis to build an \textit{inter-node} roofline model, which characterizes performance in terms of the ratio of network traffic to local work.

Assume a two-dimensional grid of matrix tiles, where the matrices A ($m \times k$), B ($k \times n$) B, and C ($m \times n$) are distributed amongst $p$ processors laid out in $\sqrt{p} \times \sqrt{p}$ tile grids.  Tile dimensions for A, B, and C are then $\frac{m}{\sqrt{p}} \times \frac{k}{\sqrt{p}}$, $\frac{k}{\sqrt{p}} \times \frac{n}{\sqrt{p}}$, and $\frac{m}{\sqrt{p}} \times \frac{n}{\sqrt{p}}$ respectively, with $\sqrt{p}$ tile columns and $\sqrt{p}$ tile rows in each matrix.  For a dense matrix, the total number of elements in each tile is the product of its dimensions, while for a sparse matrix we approximate the number of nonzeros per tile as the matrix density $d$ multiplied by the product of the tile dimensions.

% TODO: get plot of this processor grid.

For our stationary C RDMA-based algorithm for SpMM, in each iteration, each process will issue two RDMA get operations, one to retrieve a sparse tile of $A$, and one to retrieve a dense tile of $B$.  Therefore, in each iteration, each process will retrieve exactly $\frac{kn}{p}$ elements in the dense tile of $B$, along with $\frac{dmk}{p}$ elements in a CSR data structure, for a total communication cost of $\frac{kn}{p} + 2\frac{dmk}{p} + \frac{m}{\sqrt{p}} + 1$ elements.

% TODO: note somewhere that this holds for stationary A/B

If we offset the order of iteration of the $k$ loop by $i + j$, as discussed in Section~\ref{sec:optimizations}, we can also ensure that communication will be balanced with each process sending exactly two tiles per iteration, one of the A matrix, and one of the B matrix.
To prove this, assume tile $(i, j)$ is stored on processor $i(N + 1) + j$, where $i \in [0,N)$, $j \in [0,N)$, and $N = \sqrt{p}$.
% TODO: remove this proof?
\begin{proof}
Towards a contradiction, assume that $i(N + 1) + j$ is not unique.  Then, $i(N + 1) + j = i'(N + 1)+ j'$.  Assume $i' > i$.  Then $j = (i' - i)(N + 1) + j'$.  Since $i' - i \geq 1$, $j > N + 1$, which is a contradiction.
\end{proof}

With the offset, each process sends and receives exactly $\frac{kn}{p} + 2\frac{dmk}{p} + \frac{m}{\sqrt{p}} + 1$ elements over the network in each iteration.  In a fully connected network, each of these transfers is independent and will not interfere with each other.  Many modern datacenter networks, such as that used by Summit, are indeed fully connected fat trees where this assumption holds completely.  Many other network topologies, such as Dragonfly, approximate a fully connected network.

% In the case of Cannon's algorithm, which reaches the theoretical lower bound for distributed matrix multiplication without replication, each process also sends $2(N / \sqrt{p})^2$ elements in each iteration, which includes one tile of A, continually passed on to the right, and one tile of B, continuously passed down.  The popular SUMMA algorithm also achieves within a log factor of this lower bound.

% TODO: cite Cannon

Next, we present an \textit{inter-node} roofline model that predicts the performance of each iteration of the distributed matrix multiplication.  The roofline model predicts the maximum possible performance of a computational kernel depending on its \textit{arithmetic intensity}, or number of operations performed per byte that must be fetched from memory.  Typically, the roofline model is used to characterize serial or multi-threaded compute kernels in terms of how compute or bandwidth intensive they are, where the limit on ``compute'' is defined by the processor's arithmetic peak, and ``bandwidth'' is determined by memory bandwidth.  Here, we develop an \textit{inter-node} roofline model to characterize performance of our distributed memory, RDMA-based matrix multiply algorithms.  In this instance, the bandwidth involved is network communication, and the roofline peak of local operations serves as our ``arithmetic peak.''.  In our \textit{inter-node} roofline model, we compute the inter-node arithmetic intensity of each iteration as the number of flops performed divided by the number of bytes that must be communicated over the network.  The flat portion, providing the ``roof'' on performance of our inter-node roofline, is the \textit{local} roofline peak for our local SpMM or SpGEMM calls.

We first build a regular \textit{local} roofline model to characterize the performance of our local matrix multiplies.  For local SpMM and SpGEMMs, precise roofline models can be difficult to compute, often requiring analysis of the actual matrices themselves, since they depend not only on the size of the matrices and the number of nonzeros, but also the sparsity patterns of the matrices involved.  For SpMM, we can calculate an approximate upper bound on arithmetic intensity as the number of flops to be performed in the SpMM divided by the total size of the sparse and dense matrices in bytes.  This upper bound assumes perfect cache performance for B and C, and depending on the nonzero pattern, values may have to be reloaded from memory, resulting in lower performance.  For a local SpMM operation multiplying a tile of A times B, with the tile sizes derived above, and adding the variable $w$, which is the number of bytes per word, we compute the arithmetic intensity of our local SpMM operations.
\\
\textit{local SpMM AI}
\scalebox{1.0}
{
  $\displaystyle= \frac{2(\frac{dmk}{p})(\frac{n}{\sqrt{p}})}{w(2\frac{dmk}{p} + \frac{m}{\sqrt{p}} + 1 + \frac{mn}{p} + \frac{kn}{p})}$
}
\\\\
That is, the number of flops to be performed, divided by the total number of bytes in A, B, and C.  To calculate the local roofline peak, which is the ``roof'' on our inter-node roofline model, we multiply this arithmetic intensity by the memory bandwidth of the local processor $B$, followed by a max operation with the arithmetic peak.

To compute the arithmetic intensity of an iteration of our distributed SpMM algorithm using the \textit{inter-node} roofline model, we divide the number of flops performed, which is the same expression from the numerator of the local SpMM arithmetic intensity, by the total number of bytes to be communicated over the network, equal to the total number of bytes in the tiles of A and B.
\\
\textit{inter-node SpMM AI}
\scalebox{1.0}
{
  $\displaystyle= \dfrac{2(\frac{dmk}{p})(\frac{n}{\sqrt{p}})}{w(2\frac{dmk}{p} + \frac{m}{\sqrt{p}} + 1 + \frac{kn}{p})}$
}
\\\\
We can then use this roofline to characterize the performance of SpMM on the Summit supercomputer.  Figure~\ref{fig:spmm_roofline} shows the roofline plot for our inter-node SpMM roofline model, using the isolates subgraph2 sparse matrix, and evaluating performance for various numbers of columns in the dense matrix.  In the plot, the blue sloping portion represents the region that is bound by network communication, and as such the bandwidth-bound portion has a slope of 3.83~GB/s, which is each GPU's share of injection bandwidth on Summit.  The brown roofline at the top is the 32-bit floating point arithmetic peak, 16 TFlops/s for an Nvidia Tesla V100 GPU.  Each of the other dotted horizontal lines represents a \textit{local} roofline peak, which in the inter-node roofline model replaces the arithmetic peak.  The vertical solid lines represent the roofline peak for a particular problem size, while the dots indicate achieved performance of our RDMA algorithm.  Since all three of the problem sizes plotted are well into the bandwidth-bound portion of the inter-node roofline plot, we expect them to be bound by network communication.  The SpMM roofline model was across 24 GPUs.  Note that the achieved performance slightly exceeds the roofline bound based on newtork bandwidth.  This is because of the high-speed NVLink interconnect providing higher bandwidth for intra-node transfers.

\begin{figure}
  \centering
  \includegraphics[width=0.49\linewidth]{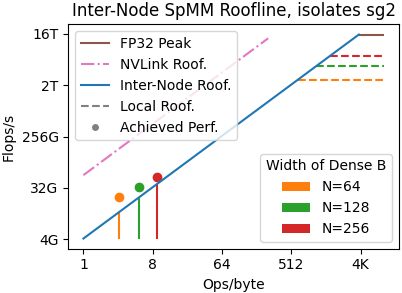}
  \includegraphics[width=0.49\linewidth]{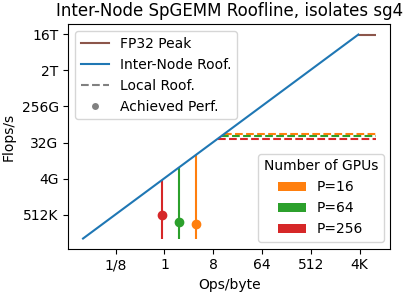}\\
  \caption{\edit \textit{Inter-node} roofline plots for SpMM and SpGEMM with a 2D distribution.  SpMM plot models performance for different widths of the dense B matrix at a fixed scale (24 GPUs), while SpGEMM models performance at different scales.  Dashed horizontal lines represent \textit{local roofline peaks} for SpMM and SpGEMM operations, while vertical lines represent \textit{inter-node} roofline peaks for particular problems.}
  % TODO: remove some datasets, reduce plot size.
  %       put x-axis labels on every graph, fix y-axis limits
\label{fig:spmm_roofline}
  % \vspace{-1.5em}
\end{figure}

While we can also construct an inter-node roofline model for SpGEMM, it is not possible to write a general formula for FLOPs performed for a particular matrix size and density.  This is because in SpGEMM, the number of FLOPs performed depends on the particular sparsity patterns of the matrices, and as such the number of FLOPs can vary significantly for matrices with the same dimensions and number of nonzeros.  In light of this, we use the function $\text{FLOPS}(A, B)$ to represent FLOPs performed, and in our roofline plot we use average FLOP values calculated experimentally.  As such, we can represent the inter-node roofline model's arithmetic intensity as the number of flops performed divided by the size of $A$ and $B$.
\\
\textit{inter-node SpGEMM AI}
\scalebox{0.85}
{
  $\displaystyle = \dfrac{\text{FLOPS}(A, B)}{w(\frac{2dmk}{p} + \frac{m}{\sqrt{p}} + 1 + \frac{2dkn}{p} + \frac{k}{\sqrt{p}} + 1)}$
}
\\\\
For our local roofline model, we use the bound on local SpGEMM arithmetic intensity by Gu, \textit{et al}., which is representative of modern SpGEMM algorithms and expresses arithmetic intensity in terms of the compression factor $cf$, the number of flops performed per nonzero output, and the number of bytes to express each nonzero $b$.
\\
\textit{local SpGEMM AI}
{
  $\displaystyle = \frac{cf}{(3 + 2cf)*b}$
}
\\\\
Note that the roofline model is dependent on the sparsity structure, since $cf$ depends on the particular sparse matrices being multiplied.
Our inter-node SpGEMM roofline model is plotted in Figure~\ref{fig:spmm_roofline}.  To obtain realistic values for \textit{cf} and the number of flops performed \textit{\text{FLOPS}(A, B)}, we performed each of the component local SpGEMM operations in the distributed SpGEMM operation for the isolates subgraph4 matrix, recording their values for \textit{cf} and \textit{\text{FLOPS}(A, B)} to compute average values for different numbers of GPUs $P$.  In the SPGEMM plot, inter-node roofline peaks are much closer to their local roofline peaks than in the SpMM plot.  Thus, our roofline model suggests SpGEMM is significantly less network communication-bound than SpMM, although still communication bound.  {\edit We discuss more conclusions from our roofline model in Section~\ref{sec:eval}.}

% TODO: cite Ariful's work, bit about how rooflines work.

% \vspace{-0.5em}
\section{Implementation}
\subsection{Communication Layer}
As discussed in Section~\ref{sec:rdma}, the current generation of GPU-based supercomputers offers GPUDirect RDMA over Infiniband networks, which ensures that issuing one-sided operations will be efficiently executed in hardware. In GPUDirect RDMA, the CPU prepares an Infiniband request, which it sends to the local network interface card (NIC).  When the NIC receives the request, it will copy the data directly over the network from GPU to GPU, without staging data through the CPU.  When copies take place between GPUs within the same node, the high-bandwidth NVLink fabric is used.

All of our RDMA-based implementations use NVSHMEM for communicating data between GPUs.  NVSHMEM uses GPUDirect RDMA over Infiniband to transfer data when the GPUs involved are on separate nodes, and NVLink to transfer data when the GPUs involved are within the same node.

\subsection{Data Structures}
We implement sparse and dense distributed matrix data structures using BCL, an RDMA-based distributed data structures library in written in C++.  Both dense and sparse matrix data structures split the distributed matrix up into evenly sized tiles, which are then assigned to processors using a processor grid, as in ScaLAPACK.
% TODO: citation to ScaLAPACK.
As discussed in Section~\ref{sec:data_structures}, each process has a copy of a directory of remote pointers, which point to the matrix tiles stored in pinned memory, directly accessible through RDMA operations using NVSHMEM.  In the dense case, a single remote pointer points to a dense tile data structure, and in the sparse case three remote pointers point to the values, row pointer, and column index arrays of a compressed sparse row (CSR) data structure.  All data is stored directly on GPUs, and can be copied directly from GPU to GPU over the network using BCL's NVSHMEM backend.

% \vspace{-0.5em}
\subsection{Data Access Primitives}
In terms of data access primitives, all direct manipulation of data is performed using put and get RDMA operations.  To retrieve a tile of the dense matrix, a process will issue an RDMA get operation to retrieve all or part of a remote tile.  Similarly, to write to the tile, a process can issue a remote put operation.  In the sparse case, reading from a tile involves issuing remote get operations to retrieve each of the value, row pointer, and column index arrays in a CSR data structure.  Due to the complexities involved in modifying a CSR data structure in place, only the process that owns a sparse tile is permitted to modify it.  This is done by calling a function \texttt{replace\_tile()} to replace the old tile with a new one, followed by a collective function \texttt{renew\_tiles()} that will make all sparse tile modifications visible to other processes.  As discussed in Section~\ref{sec:modify_remote}, when a process needs to send an update to a remote tile, as in the A and B stationary algorithms, we use a distributed queue to send a pointer to the update to the process that owns a particular tile, who will then perform an accumulation at the destination tile.  For the update queue, we use BCL's \texttt{CheckSumQueue}, which enqueues a pointer using a single fetch-and-add operation and an RDMA put, while allowing simultaneous enqueues and dequeues.

As discussed in Section~\ref{sec:reading_tiles}, the primary primitive for accessing remote tiles of the matrix is \texttt{get\_tile()}, which fetches a remote tile of the matrix, and has the same API for both dense and sparse matrices.  In the asynchronous version, we return a future object, which will return the local object when the method \texttt{get()} is called.  In order to ensure the best possible performance and simplify memory management, we allocate most of each GPU's memory as shared memory that can be addressed remotely by NVSHMEM, and use our own memory allocator to allocate memory when necessary.  This (1) ensures that the local destination buffers in remote get operations are allocated in the shared memory segment, which is a requirement for NVSHMEM transfers over Infiniband, and (2) reduces the overhead of memory allocation during the algorithm, since \texttt{cudaMalloc} tends to be much less efficient than using a custom memory allocator.

% \vspace{-0.5em}
\subsection{MPI SUMMA Implementations}
For one comparison benchmark, we also implement the bulk synchronous SUMMA algorithm using CUDA-aware MPI.  In the MPI implementations, we use \texttt{MPI\_Bcast} to broadcast tiles of the matrix using row and column communicators.  We use the CUDA-aware API in IBM Spectrum MPI to perform our broadcasts directly on data that sits on the GPU, and IBM Spectrum MPI is configured to take advantage of GPUDirect RDMA to copy data directly from GPU-to-GPU over the Infiniband network.  Note that the MPI SUMMA implementation only runs on square processor grids, so we run this implementation on perfect square numbers of processors.

% TODO: cite some Summit documentation?
% \vspace{-0.5em}
\section{Evaluation}
\label{sec:eval}
We evaluated our RDMA-based sparse matrix multiply implementations in both single-node and multi-node environments.
Multi-node experiments are run on the Summit supercomputer at the Oak Ridge Leadership Computing Facility (OLCF) at Oak Ridge National Laboratory.  A Summit node has 6 Nvidia Tesla V100 GPUs, each with 16 GB of memory, and the nodes are connected with dual-rail EDR InfiniBand with a link bandwidth of 23~GB/s.  Within a node, the GPUs are connected with Nvidia's NVLink interconnect.  Single-node experiments are run on a DGX-2 system that is part of the Bridges-2 system at the Pittsburgh Supercomputing Center.  The DGX-2 has 16 Nvidia Tesla V100 GPUs with 32 GB of memory, which are fully connected by Nvidia's NVLink interconnect.  Both systems use NVLink 3.0, which provides 50 GB/s of link bandwidth.  We ran experiments for matrices whose single-GPU runtime took less than a second in the single-node DGX-2 environment, while we used the multi-node environment for matrices which took longer.

On Summit, all codes were compiled with GCC 8.1.1, CUDA 11.2.0, and IBM Spectrum MPI 10.3.1.2.  On the single-node DGX-2 system, codes were compiled with GCC 10.2.0, CUDA 11.1.0, and Open MPI 4.0.5.  Our RDMA-based implementations use NVSHMEM 2.0.2.  All performance experiments use CuSPARSE for local matrix computation.  {\edit In addition to our own SpMM and SpGEMM implementations, we also benchmark CombBLAS's GPU SpMM and PETSc's GPU SpGEMM~\cite{bulucc2011combinatorial,petsc-web-page}. We corresponded with CombBLAS's authors on how to build and execute the \texttt{gpu} branch of CombBLAS on Summit, and with OLCF staff for how to properly build PETSc with GPU support.}

{\edit The sparse matrices in our experiments are detailed in Table~\ref{table:spmm_details}. Load imbalance is defined as \textit{max}(\textrm{nnz}($A_{i,j}$))/\textit{ave}(\textrm{nnz}($A_{i,j}$)) where $A_{i,j}$ is the matrix assigned to a single processor $P(i,j)$. We multiply these matrices against dense matrices with 128 and 512 columns. All matrices use 32-bit floating point values, and for indices, we use 32-bit integers except on "Isolates, Subgraph2" and "Friendster," which require 64-bit indices due to their size. }

% TODO: list CUDA, MPI, OpenSHMEM, versions.

% TODO: cite Summit documentation

\begin{figure}
  \centering
  \includegraphics[width=\linewidth]{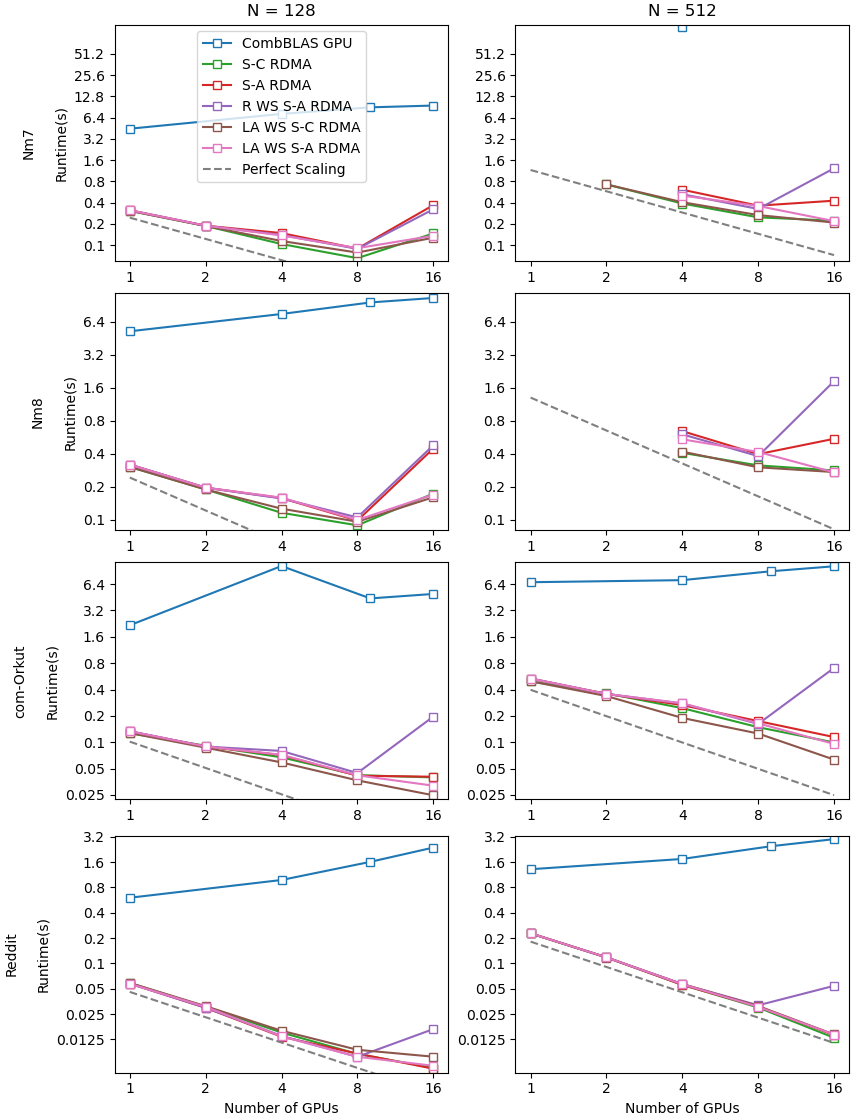}\\
  % \vspace{-0.5em}
  \caption{\edit Single-node runtimes for SpMM, with different numbers of columns $N$ in the dense matrix B.}
  % TODO: remove some datasets, reduce plot size.
  %       put x-axis labels on every graph, fix y-axis limits
\label{fig:dgx_spmm}
\end{figure}

\begin{table}[t]
    \centering
    \caption{Matrices used in our experiments. ``load imb.'' lists the imbalance in number of nonzeros when split amongst 100 processes on a $10\times 10$ {\edit 2D process grid.} }
    % \scalebox{0.90}
    \resizebox{\columnwidth}{!}
    {
        \begin{tabular}{l l r r r r}
        \toprule
        Sparse matrix ($A$) & kind & $m=k$ & $\mathrm{nnz}(A)$ & load imb. \\
        \toprule
        Mouse Gene & Biology & 45.1K & 29.0M & 2.13 \\
        ldoor & Structural & 952K & 46.5M & 8.23 \\
        reddit & GNN & 233K & 115M & 1.08 \\
        nlpkkt160 & NLP & 8.3M & 230M & 9.46 \\
        Amazon Large & GNN & 14.3M & 230M & 3.78 \\
        com-Orkut & NMF & 3.1M & 234M & 8.15 \\
        Nm7 & Eigen & 5.0M & 648M & 6.38 \\
        Nm8 & Eigen & 7.6M & 592M & 6.48 \\
        Isolates, Subgraph4 & Biology & 4.4M & 327M & 1.00 \\
        Isolates, Subgraph2 & Biology & 17.5M & 5.2B & 1.00 \\
        Friendster & Graph & 62.5M & 3.4B & 7.68\\
        \bottomrule
        \end{tabular}
    }
    \label{table:spmm_details}
\end{table}

% TODO: add all SpGEMM matrices to table

% \vspace{-0.5em}
\subsection{SpMM Experiments}
\label{sec:spmm_eval}
To evaluate our RDMA-based algorithms for sparse times dense matrix multiplication (SpMM), we performed strong scaling experiments using a number of different sparse matrices multiplied by dense matrices of different widths.  

Figures~\ref{fig:dgx_spmm} and \ref{fig:spmm_times} show the results for a variety of different algorithms.  The RDMA-based algorithms include a stationary C algorithm (``S-C RDMA''), a stationary A algorithm (``S-A RDMA''), stationary A algorithm with random workstealing (``R WS S-A RDMA''), and locality-aware workstealing with stationary A and C distributions (``LA WS S-C RDMA'' and ``LA WS S-A RDMA'').  Bulk synchronous benchmarks include our own CUDA-aware MPI SUMMA implementation (``BS SUMMA MPI''), and {\edit CombBLAS's GPU SpGEMM kernel (``CombBLAS GPU'')}.  We use matrix sizes representing typical workloads used in a range of real-world applications, such as Graph Neural Networks (GNNs), iterative methods, and graph algorithms\cite{huang2020ge,tripathy2020reducing,hu2020featgraph}.

The implementations generally scale well up to at least 8 GPUs on the single-node DGX-2 system in Figure~\ref{fig:dgx_spmm}, where each GPU has 50 GB/s of bandwidth.  Some versions suffer a slowdown at 16 GPUs, especially on the smaller Nm7 and Nm8 matrices, where at 128 columns of B there is insufficient work to keep the GPUs fully occupied and still amortize data movement. This is particularly true for the stationary A algorithms because of an increase in the number of accumulation operations that become necessary as the number of tiles, and thus the number of local matrix multiply operations, increases.  

For the larger matrices run on distributed memory in Figure~\ref{fig:spmm_times}, the shared links give each GPU only 3.8 GB/s on average. This significantly limits scalability, especially for the communication-bound case with a smaller dense B matrix.  However the RDMA-based implementations outperform the bulk synchronous implementations for most of these configurations, and that benefit  is greatest when the number of columns in the dense matrix B is lowest.  This is likely because these problems have less computation and are therefore more sensitive to communication cost, both message latency and load imbalance from communicating unequal size blocks of A.  A larger B matrix increases the both computation time and time for well-balanced communication of B.  The stationary A algorithms have the same issue of increased accumulation cost as on the single node system, but due to the communication-bound nature of the multi-node problems, the overall effect is not as pronounced.

\begin{figure}
  \centering
  \includegraphics[width=\linewidth]{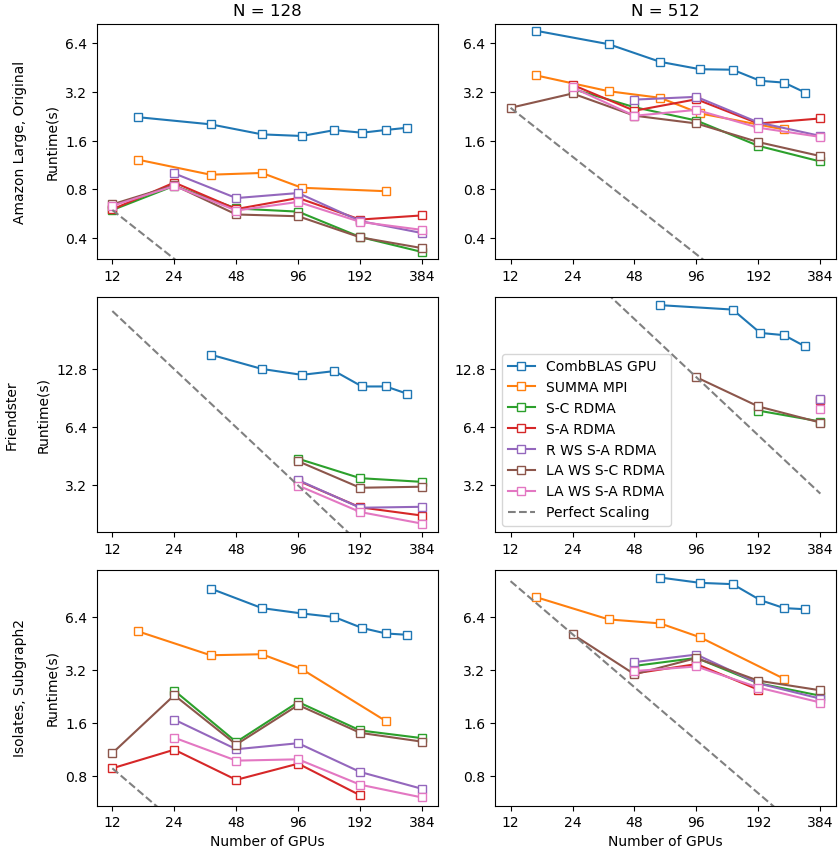}\\
  \caption{\edit Multi-node runtimes for SpMM, with different numbers of columns $N$ in the dense matrix B.}
  % TODO: remove some datasets, reduce plot size.
  %       put x-axis labels on every graph, fix y-axis limits
\label{fig:spmm_times}
  % \vspace{-1em}
\end{figure}

The multi-node scaling results are also influenced by the fact that some transfers happen over the local NVLink fabric, which is much faster than Infiniband.  As the number of GPUs increases, the number of transfers that occur over local NVLink fabric decreases, and the lower bandwidth between nodes plays a larger role.

Scaling performance is best on larger matrices, such as isolates, and scaling is also better with wider B matrices.  The reasons for this are likely twofold: (1) there is simply more work to be done with a larger dense matrix, and so scaling performance is better, and (2), as demonstrated by our inter-node roofline performance model in Figure~\ref{fig:spmm_roofline}, the wider the B matrix, the more arithmetically intense the operation becomes, and the less bound by network communication.

% TODO: say more about scaling?

The relative performance of the RDMA SpMM algorithms generally depends on the relative sizes of the A, B, and C matrices.  If one of the matrices is particularly large, then it is most efficient to keep that matrix stationary while communicating the two other, smaller matrices.  When the sparse matrix is not too large in relation to the sizes of the dense matrices, stationary C performs best, which follows from the fact that it performs accumulations locally, removing the need for remote queues, which add some additional overhead.  For problems where the sparse matrix is sufficiently larger than the dense matrices, the stationary A algorithm performs best.  This trend can be seen in the multi-node plots for the matrix friendster, where stationary A performs better for $N=128$ (meaning the dense B matrix is small), but that is reversed for $N=512$ (meaning the dense B becomes bigger).  As with most real-world graph applications, all of our sparse matrices are square, which means that the B and C matrices are the same size.  This means there is no benefit in terms of communication volume to leaving B stationary instead of C, even while stationary B adds additional overhead due to remote queues.  As such, we do not explore a stationary B implementation for SpMM.

Our RDMA-based workstealing algorithms are able to achieve performance improvements for some of the less load balanced matrices, such as com-Orkut and friendster, particularly at higher node counts, where the smaller amount of work per node is more likely to lead to some nodes being left with uneven amounts of work.  The locality-aware workstealing algorithms tend to perform much better than the random workstealing algorithm, which often incurs a high cost when performing workstealing, since it is not guaranteed that any of the involved matrix tiles will be local.  In Table~\ref{table:components_spmm}, we can see that the locality-aware workstealing algorithm is able to reduce the amount of time lost due to load imbalance compared to the other RDMA implementations.
It should be noted that load imbalance numbers for the RDMA implementations cannot be directly compared to the bulk synchronous MPI implementation, since time lost to load imbalance is tied up in MPI's broadcast operations, which perform communication, but also enforce synchronization.

% As shown in Figure~\ref{fig:spmm_components}, we do see a modest but measurable decrease in the amount of time wasted due to load imbalance (``Barrier'') when comparing the workstealing stationary A algorithm to its non-workstealing counterpart.

% Looking at the bulk synchronous SUMMA implementation, we see that the vast majority of time is taken up with ``Sync,'' which for the MPI-based implementation is monitoring time spent performing \texttt{MPI\_Bcast} operations to communicate the data.  It should be noted that although the time for ``Barrier'' is quite low in the SUMMA implementation, this does not mean that there is no time lost to load imbalance in the MPI implementation.  This is because the \texttt{MPI\_Bcast} operations include an implicit barrier operation, forcing processes to wait if there is an imbalance in computation or communication.

%\begin{figure*}
%  \centering
%  % \includegraphics[width=\linewidth]{plots/spmm_times.pdf}\\
%  % \includegraphics[width=\linewidth]{plots/spmm_components.png}\\
%  TODO: Add graph with SpMM components
%  \caption{{\edit Runtime component breakdowns for SpMM, $N = 256$ columns in the dense matrix B.}}
%  % TODO: remove some datasets, reduce plot size.
%  %       put x-axis labels on every graph, fix y-axis limits
%\label{fig:spmm_components}
%\end{figure*}

% \vspace{-0.5em}
\subsection{SpGEMM Experiments}
To evaluate our RDMA-based SpGEMM algorithm, we perform the computation $C = AA$ for a number of different sparse matrices, representative of a number of different graph algorithms, such as Markov clustering~\cite{vandongen00}, triangle counting~\cite{azad2015parallel}, and transitive reduction~\cite{guidi2020parallel}.  {\edit We again benchmark RDMA-based stationary C and A algorithms, along with a work-stealing version, a SUMMA implementation in MPI, and PETSc's SpGEMM using CuSPARSE.}

% TODO: get results for at least an MPI SUMMA version

Strong scaling performance results are shown in Figure~\ref{fig:spgemm_times}.
Unlike SpMM, both the single- and multi-node environments are more likely to be compute bound. As a result, all the implementations have similar performance, because they are all performing the same computations.  The only exception is PETSc, which is significantly slower, probably because it is not utilizing GPUDirect RDMA.  On the large isolates matrix, our bulk synchronous CUDA-aware MPI implementation matches that of the RDMA implementation.  On mouse gene, which is somewhat smaller and less load-balanced, we observe good scaling until very high concurrencies, but the RDMA-based implementations achieve higher overall performance as well as somewhat better scaling.  On nlpkkt160, which has an uneven distribution of nonzeros, we observe poor performance in the multi-node experiments due to imbalance in both communication and computation.  However, the RDMA implementations still perform better than the bulk synchronous implementations.  Unlike SpMM, we do not meet our SpGEMM model's roofline bound on Summit, as shown in Figure~\ref{fig:spmm_roofline}.  This is because the \textit{local} cuSPARSE operations are not meeting the local roofline bound, evidenced by SpGEMM being compute bound in Table~\ref{table:components_spgemm}.  This likely indicates that there are opportunities for optimization in the local cuSPARSE SpGEMM operations.

For single-node experiments, we achieve better scaling on the highly load imbalanced matrices nlpkkt160 and ldoor, likely because communication imbalance is eliminated due to by the significant increase in bandwidth, along with the fact that at lower concurrencies each GPU is left with a larger portion of the matrix, reducing the likelihood of load imbalance.

\begin{table}[t]
    \centering
    \caption{Component breakdown for selected matrices.}
    \begin{subtable}{0.5\textwidth}
      % \vspace{-0.5em}
    \caption{Component breakdown for SpMM, N = 256 columns.}
    % \scalebox{0.79}
    \resizebox{\textwidth}{!}
    {
        \begin{tabular}{l l l r r r r r}
        \toprule
          Env. & Matrix & Alg. & \#GPUs & Comp. & Comm. & Acc. & Load Imb.\\
          \hline
          Summit & Amazon & S-C & 24  & 0.02 & 1.3 & - & 0.4\\
                 &        &     & 192 & \textasciitilde 0 & 0.5 & - & 0.2\\
                 &        & S-A & 24  & 0.02 & 1.1 & 0.4 & 0.3\\
                 &        &     & 192 & \textasciitilde 0 & 0.6 & 0.2 & 0.2\\
                 &        & S-C LW & 24 & 0.01 & 1.2 & 0.03 & 0.3\\
                 &        &        & 192 & \textasciitilde 0 & 0.6 & 0.01 & 0.1\\
                 &        & MPI    & 16 & 0.02 & 2.1 & - & -\\
                 &        &        & 256 & ~0 & 1.0 & - & -\\
          \hline
          DGX-2 & Nm-7 & S-C & 4  & 0.44 & 0.01 & 0.02 & 0.14\\
                &      & S-C & 16 & 0.19 & \textasciitilde 0 & 0.09 & 0.10\\
                &      & S-A & 4  & 0.42 & \textasciitilde 0 & 0.07 & 0.12\\
                &      & S-A & 16 & 0.09 & 0.00 & 0.02 & 0.10\\
                % &      & S-C LW & 4  & 0.15 & 0.03 & \textasciitilde 0 & 0.04\\
                % &      & S-C LW & 16 & 0.06 & 0.02 & 0.003 & 0.06\\
                 &        & MPI    & 16 & 0.64 & 0.51 & 0.01 & 0.15\\
                 &        &        & 256 & 0.08 & 0.15 & 0.01 & 0.08\\
        \bottomrule
        \end{tabular}
    }
    \label{table:components_spmm}
\end{subtable}
  \begin{subtable}{0.5\textwidth}
    \caption{Component breakdown for SpGEMM.}
    % \scalebox{0.75}
    \resizebox{\textwidth}{!}
    {
        \begin{tabular}{l l l r r r r r}
        \toprule
          Env. & Matrix & Alg. & \#GPUs & Comp. & Comm. & Acc. & Load Imb.\\
          \hline
          Summit & Mouse Gene & S-C & 24  & 0.02 & 1.3 & - & 0.4\\
                 &        &     & 192 & \textasciitilde 0 & 0.5 & - & 0.2\\
                 &        & S-A & 24  & 0.02 & 1.1 & 0.4 & 0.3\\
                 &        &     & 192 & \textasciitilde 0 & 0.6 & 0.2 & 0.2\\
                 &        & S-C LW & 24 & 0.01 & 1.2 & 0.03 & 0.3\\
                 &        &        & 192 & \textasciitilde 0 & 0.6 & 0.01 & 0.1\\
                 &        & MPI    & 16 & 0.02 & 2.1 & - & -\\
                 &        &        & 256 & ~0 & 1.0 & - & -\\
          \hline
          DGX-2 & Mouse Gene & S-C & 4  & 2.14 & \textasciitilde 0 & 0.02 & 0.72\\
                &      &           & 16 & 0.55 & \textasciitilde 0 & 0.01 & 0.21\\

                &      & S-A & 4  & 2.14 & \textasciitilde 0 & 0.38 & 0.49\\
                &      &     & 16 & 0.55 & \textasciitilde 0 & 0.11 & 0.13\\

                % &      & S-C LW & 4  & 0.15 & 0.03 & \textasciitilde 0 & 0.04\\
                % &      & S-C LW & 16 & 0.06 & 0.02 & 0.003 & 0.06\\
        \bottomrule
        \end{tabular}
    }
    \label{table:components_spgemm}
  \end{subtable}
\end{table}

\begin{figure}[thb]
 \centering
  \subcaptionbox{Single-node experiments.}
 {%
 \includegraphics[width=.492\linewidth]{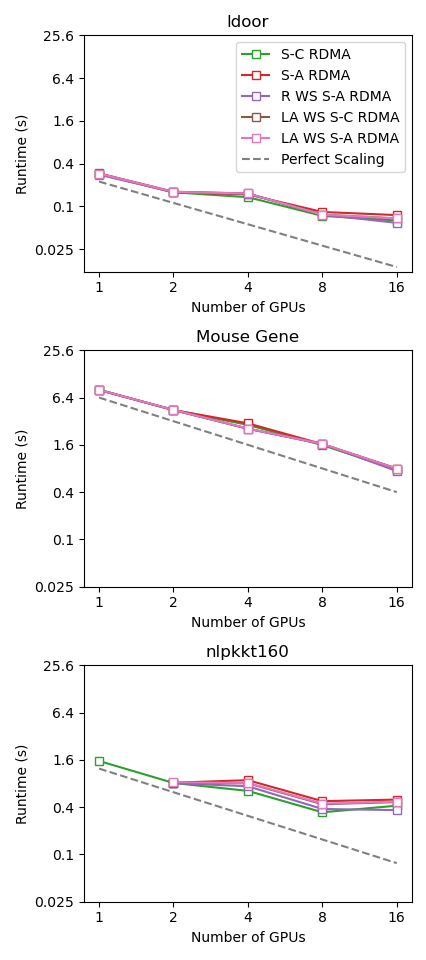}%
 }
 \hfill
  \subcaptionbox{Multi-node experiments.}
 {%
 \includegraphics[width=.492\linewidth]{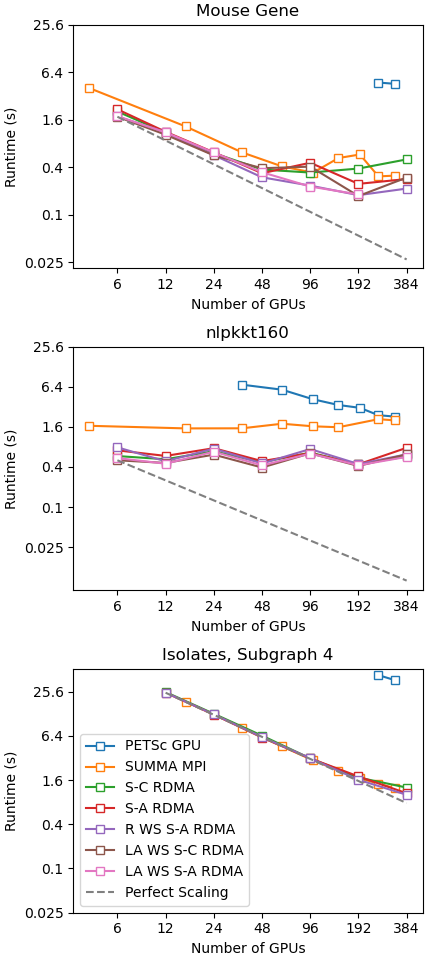}%
 }
 \caption{SpGEMM strong scaling experiments.}
 \label{fig:spgemm_times}
\end{figure}

%\begin{figure*}
%  \begin{subfigure}{0.51\textwidth}
%  \includegraphics[width=\linewidth]{plots/dgx_spgemm_cut_new.png}\\
%    \caption{Single-node SpGEMM experiments.}
%  \end{subfigure}
%  \begin{subfigure}{0.51\textwidth}
%  \includegraphics[width=\linewidth]{plots/summit_spgemm_cut.png}\\
%    \caption{Multi-node SpGEMM experiments.}
%  \end{subfigure}
%\end{figure*}

%\begin{figure*}
%  \centering
%  % \includegraphics[width=\linewidth]{plots/spmm_times.pdf}\\
%  TODO: add graph with SpGEMM components
%  % \includegraphics[width=\linewidth]{plots/spgemm_components.png}\\
%  \caption{\edit Runtime component breakdowns for SpGEMM, multiplying each matrix by itself.}
%  % TODO: remove some datasets, reduce plot size.
%  % \vspace{-1em}
%\label{fig:spgemm_components}
%\end{figure*}

% \vspace{-0.5em}
\section{Related Work and Conclusions}
The C++ PGAS library DASH, which uses RDMA through MPI's one-sided communication API, includes dense matrix and vector data structures as well as an implementation of stationary C dense matrix multiplication \cite{Fuerlinger:2016:DASH,Moessbauer:CoArray}.  However, DASH does not include any support for sparse matrices or GPUs.

The Combinatorial BLAS library (CombBLAS)~\cite{azad2021combinatorial} is an MPI-based library that provides distributed data structures for dense and sparse matrices, also implementing a number of distributed sparse matrix multiply algorithms. While focusing more on tensor computations, Cyclops Tensor Framework (CTF)~\cite{solomonik2013cyclops} also supports several distributed sparse matrix multiplication algorithms.

Both CombBLAS and CTF implement also include so-called communication-avoiding algorithms for sparse matrices, which replicate some of the matrices to reduce communication costs. These ideas have also been applied to SpGEMM~\cite{azad2016exploiting} and SpMM within the context of graph neural network training~\cite{tripathy2020reducing}. Our work on asynchronous, RDMA-based implementations is orthogonal to these techniques, and could be combined.

% Distributed matrix-matrix multiplication has seen a sustained interest among the HPC community. One of the most popular developments has been due to communication-avoiding (CA) algorithms~\cite{ballard2011minimizing} that replicate input, output, or intermediate data to asymptotically reduce communication. CA algorithms have been applied to SpGEMM~\cite{azad2016exploiting} and SpMM within the context of graph neural network training~\cite{tripathy2020reducing}. Both CombBLAS and CTF implement CA algorithms for sparse matrices. Our work on asynchronous, RDMA-based implementations is orthogonal to CA algorithms, as our approach can be applied to 3D algorithms. 

This work explores a large space of possible communication and load balancing optimization for two sparse matrix operations, which are key to many simulation, analysis, and learning applications.  We demonstrate that RDMA based communication can have a significant advantage over global collectives for some problems, and that dynamic work stealing can also contribute to improved scaling. Our extensive set of experiments on multiple matrices and two different machine configurations also highlight the importance of having high speed networking hardware that is balanced with the computational capabilities for this class of algorithms.  Our inter-node roofline model provides a useful tool for evaluating performance and identifying opportunities for optimization.

\begin{acks}
This work was supported in part by the National Science Foundation Graduate Research Fellowship Program under Grant No. GE 1752814 as well as
under National Science Foundation Award No. CCF-1823034.
It was also funded in part by the Advanced Scientific Computing Research (ASCR) program within the
Office of Science of the DOE under contract number
DE-AC02-05CH11231.
This work used resources of the National Energy Research Scientific Computing Center (NERSC) at Lawrence Berkeley National Laboratory supported by the Office of Science of the DOE under Contract No. DE-AC02-
05CH11231
as well as Extreme Science
and Engineering Discovery Environment (XSEDE) Bridges-2 at the Pittsburgh Supercomputing Center through allocation ASC180051.
\end{acks}

% TODO: cite SHAAD, although it's unclear exactly what way to bring it up.

% \vspace{-1em}
% \section{Conclusion}
% In this paper, we presented asynchronous RDMA-based algorithms for sparse matrix multiplication, including SpMM and SpGEMM.  We developed a roofline performance model to characterize the performance of our algorithms and evaluated the performance of our algorithms running in a distributed environment across multiple GPUs, demonstrating asynchronous algorithms perform favorably compared to traditional bulk synchronous SUMMA.

% TODO: set bibliography style
\bibliographystyle{ACM-Reference-Format}
\bibliography{bcl-refs}

\end{document}